\documentclass[twocolumn]{iopart}
\usepackage{graphicx}% Include figure files
\usepackage{iopams}
\begin{document}

\title[Effect of many-body quantum fluctuations on matrix Berry phases]{Effect of many-body quantum fluctuations on matrix Berry phases of  a two-dimensional n-type semiconductor quantum dot}
\author{S.C. Kim$^1$, Y.J. Kim$^1$, P.S. Park$^1$, N.Y. Hwang$^{1,2}$,  and S.-R. Eric
Yang$^1$}
%\footnote{corresponding author
%eyang@venus.korea.ac.kr}}
\address{$^1$ Physics Department, Korea  University, Seoul Korea 136-713}
\address{$^2$ Physics Department, University of Toronto, Toronto M5S 1A7 Ontario, Canada}
\ead{corresponding author eyang@venus.korea.ac.kr}
\begin{abstract}
In the presence of  spin-orbit coupling  and inversion symmetry of
the lateral confinement potential a single electron does not exhibit
matrix Berry phases in quasi-two-dimensional semiconductor quantum
dots. In such a system we  investigate whether many-body correlation
effects can lead to finite  matrix Berry phases. We  find  that the
transformation properties of many-electron wavefunctions under
two-dimensional inversion operation do not allow finite matrix Berry
phases. This effect is exact and is independent of the form of
electron-electron interactions. On the other hand,
quasi-two-dimensional semiconductor quantum dots with lateral
confinement potential without inversion symmetry can have  finite
matrix Berry phases.  We find that many-body  quantum fluctuations
can change matrix Berry phases significantly in such systems.
\end{abstract}
%\pacs{71.55.Eq, 71.70.Ej, 03.67.Lx, 03.67.Pp}
%\maketitle

%\date{\today}

\section{Introduction}

Electron spins in two-dimensional semiconductors may be manipulated
electrically\cite{Da,Nitta,Loss2,Sa,ras0,Kato,Sch,Ka,Sc}. It is more
challenging to control single or a few spins  coherently  in
confined nano quantum dots \cite{Han, To,Go,Le,De,Wa,Aws}. One way
to perform such a coherent  control electrically is based on  matrix
Berry phases\cite{Sha,Wil,Zan}. There are several semiconductor
nanosystems  with spin-orbit coupling terms \cite{ras,dre} that
exhibit matrix Berry phases: they include excitons\cite{Sol},  CdSe
nanocrystals\cite{Sere2},  acceptor states of p-type
semiconductors\cite{Bern}, and ring spin filters\cite{Hat}. Recently
it has been demonstrated theoretically that it is possible to
control electrically electron spins of II-VI and III-V n-type
semiconductor quantum dots\cite{yang1} and rings\cite{yang2} by
exploiting matrix Berry phases.  In these systems spin-orbit terms
are invariant under time-reversal operation \cite{val} and  the
discrete energy levels are  doubly degenerate, and these properties
are responsible for the generation of matrix Berry phases\cite{Mead}.

Coherent manipulation can be achieved by changing external
parameters adiabatically in time. According to the theory of matrix
Berry phase\cite{Wil} the groundstate of  a doubly degenerate
Hilbert subspace of these II-VI and III-V n-type semiconductor
quantum dots changes adiabatically in time as
\begin{eqnarray}
\Psi(t)=C_1(t)\Phi(t)+C_2(t)\overline{\Phi}(t),
\label{eq:time-dep-state}
\end{eqnarray}
where $\Phi(t)$ and $\overline{\Phi}(t)$ are the instantaneous {\it
degenerate} single electron or many-body eigenstates (the overline
in $\overline{\Phi}(t)$ means time-reversal state of $\Phi(t)$). The
time dependent Schr\"{o}dinger equation for the expansion
coefficients $C_1$ and $C_2$ of Eq.(\ref{eq:time-dep-state}) can  be
written as
\begin{eqnarray}
i \hbar \dot{C}_v=-\sum_w A_{v w} C_w \qquad v=1,2,
\label{eq:time_Schrod}
\end{eqnarray}
where $A_{vw}= \hbar \sum_k(A_k)_{v,w}\frac{d\lambda_k}{dt}$ and
$\lambda_k$ are the adiabatic parameters labeled by $k$. The time
evolution of the groundstate is governed by the $2\times2$
non-Abelian vector potentials (NAVPs) between the degenerate
eigenstates
\begin{eqnarray}
(A_k)_{v,w}=i\langle \Phi_v|\frac{\partial}{\partial\lambda_k
}|\Phi_w\rangle, \label{eq:intra}
\end{eqnarray}
where $\Phi_1=\Phi(t)$ and  $\Phi_2=\overline{\Phi}(t)$.

II-VI and III-V semiconductor quantum dots  usually contain several
electrons, and many-body effects may  affect the matrix Berry phase.
Formal expressions for many-body NAVPs can be derived including many
body exchange and correlation effects\cite{yang3}. Correlation
effects are taken into account by writing many body eigenstates as a
linear combination of  single Slater determinant wavefunctions. When
odd number of electrons are present the groundstates are  doubly
degenerate.  At each time instant $t$ they can be written as linear
combinations of Slater determinant states $|{\Psi}_i\rangle$:
\begin{eqnarray}
|\Phi\rangle=\sum_{i}^Mc_i|\Psi_i\rangle,\
|\overline{\Phi}\rangle=\hat{T}\Phi
=\sum_{i}^{M}d_i|{\Psi}_i\rangle, \label{eq:groundstates}
\end{eqnarray}
where $\hat{T}$ is time reversal operator and $M$ is the number of
instantaneous Slater determinant states in the linear combinations.
These states are time reversal states of each other. (We have
suppressed  $t$ in the quantities appearing in
Eq.(\ref{eq:groundstates}), and from now on we will do so unless
explicitly written). The   diagonal elements of the  NAVPs are
\begin{eqnarray}
(A_k)_{1,1}&=&i\langle \Phi|\frac{\partial}{\partial\lambda_k}|\Phi
\rangle
=i\sum_{i}c^{*}_{i}\frac{\partial c_{i}}{\partial \lambda_{k}}+\sum_{i,j}c_i^*c_j(B_k)_{i,j},
\label{eq:Diagonal}
\end{eqnarray}
where the elements of the NAVP between  Slater determinant states
are
\begin{eqnarray}
(B_k)_{i,j}=i\langle
\Psi_i|\frac{\partial}{\partial\lambda_k}|\Psi_j  \rangle.
\label{eq:Slater}
\end{eqnarray}
(The Slater determinant states in this expression can have different
total confinement  energies). It can be shown that if $(B_k)_{i,j}$
is non-zero one can find  single-electron wavefunctions $\phi_{p}$  and $\phi_{q}$
so that
\begin{eqnarray}
(B_k)_{i,j}=(a_k)_{p,q}=i\langle
\phi_{p}|\frac{\partial}{\partial\lambda_k }|\phi_{q}\rangle.
\label{eq:inter}
\end{eqnarray}
When the single electron eigenstates $\phi_{p}$  and $\phi_{q}$
belong to different energy shells $(a_k)_{p,q}$ are called  single
electron {\it inter-shell} NAVPs\cite{yang3}. The other diagonal
elements $(A_k)_{2,2}$ are given by Eq.(\ref{eq:Diagonal}) except
that  $\Phi$ is replaced by $\overline{\Phi}$. The  off-diagonal
elements are
\begin{eqnarray}
(A_k)_{1,2}&=&i\langle
\Phi|\frac{\partial}{\partial\lambda_k}|\overline{\Phi}  \rangle
=i\sum_{i}c^{*}_{i}\frac{\partial d_{i}}{\partial \lambda_{k}}+\sum_{i,j}c_i^*d_j(B_k)_{i,j},
\label{eq:OffDiagonal}
\end{eqnarray}
with  $(A_k)_{2,1}=(A_k)_{1,2}^*$. Within this approach one can use
a Hartree-Fock approximation based on single Slater determinant
groundstates, and show that fermion antisymmetry  does not change
the value  of the matrix Berry phase.

These formal results have not been applied to investigate the
interplay between matrix Berry phase and many-body correlations of
II-VI and III-V semiconductor quantum dots. For example, not much is
known about how the effects  beyond Hartree-Fock approximation,
i.e., correlation effects,  change the matrix Berry phase. The total
many-electron Hamiltonian consists of four terms: the kinetic part
$H_K$, confinement potential  $V_C$, spin-orbit terms  $H_{so}$, and
electron-electron interactions $V$:
\begin{eqnarray}
H=H_K+V_C+H_{so}+V. \label{eq:total-Ham}
\end{eqnarray}
Note that  $V_C$ {\it may or may not} be  invariant under
two-dimensional inversion operation. However, $H_{so}$ is {\it not }
invariant under two-dimensional inversion operation, and,
consequently,
 the total Hamiltonian $H$  is not invariant under two-dimensional inversion operation, irrespective of the invariance of $V_C$.
This implies that eigenstates of the total Hamiltonian $H$ are not
eigenstates of two-dimensional inversion operator. In the absence of
many-body  effects it can be shown that, when  the lateral electric
confinement potential has inversion symmetry, i.e.,  $V_C$ is
invariant under two-dimensional inversion operation, the matrix
Berry phase is absent\cite{yang1}.  This is because off-diagonal
elements of single electron {\it intra-shell} NAVPs,
Eq.(\ref{eq:intra}), may vanish. However, in the many-electron case
the NAVPs may take finite values  since they are related to the
single electron {\it inter-shell} NAVPs, which can be non-zero, as
can be seen  from Eq.(\ref{eq:inter}). It is thus unclear whether
the matrix Berry phase  remains zero or not. In addition, it is not
understood  how many-body correlation effects change quantitatively
the matrix Berry phase when a distortion potential  breaks inversion
symmetry of the lateral electric confinement potential. Such a
quantitative estimate should be valuable in understanding
experimental results of matrix Berry phases.

In order to investigate these issues we  use the formal results of
Eqs.(\ref{eq:groundstates})-(\ref{eq:OffDiagonal}). We have
investigated the effect of   many-body correlations, and have found
that they do not induce a finite matrix Berry phase when the lateral
confinement potential is invariant under two-dimensional inversion
operation. This is an {\it exact} result. The main physics is that
although there is coupling between different single electron energy
levels, many-body correlation effects cancel this coupling. On the
other hand, for  lateral confinement potentials without inversion
symmetry we find that many-body quantum fluctuations change the
matrix Berry phase. In this case it is difficult to investigate
exactly correlation effects. We have  treated them in an
approximation that includes a finite number $M$ of many-body basis
vectors, and have performed a numerical computation to estimate the
effect of quantum fluctuations on the matrix Berry phase. Our
approximate calculation shows that  the effect of quantum
fluctuations on the matrix Berry phase becomes more  significant as
the ratio between the Coulomb strength and the single electron
energy spacing increases. The main results of our investigation  may
be tested experimentally in semiconductor dots, as  we discuss in
Sec.4.

Our paper is organized as follows.   In Sec.2 we describe our model
Hamiltonian. In Sec.3 we compute the matrix Berry phase when the
lateral confinement potential is not invariant under two-dimensional
inversion operation. Discussions are given in Sec.4.

\section{Model Hamiltonian}

The total single electron Hamiltonian of  a II-VI or III-V n-type
semiconductor quantum dot contains an electric confinement potential
and spin orbit coupling terms. An electron with the effective mass
$m^*$ of such a system can be described by the
Hamiltonian\cite{yang1}
\begin{eqnarray}
h_S&=&h_\mathrm{K}+h_\mathrm{R},\nonumber\\
h_K&=&-\frac{\hbar^2\nabla^2}{2m^*}+U(\vec{r})+V(z),\nonumber \\
h_\mathrm{R}&=&c_\mathrm{R} \left( \sigma_x k_y -\sigma_y k_x
\right), \label{eq:singleHam}
\end{eqnarray}
where the  two-dimensional lateral confinement potential is
\begin{eqnarray}
U(\vec{r})&=&\frac{1}{2}m^*\omega^2_x x^2+\frac{1}{2}m^*\omega^2_y
y^2 +V_p(x,y),
\end{eqnarray}
and the vertical  confinement potential is $V(z)$ (Here the
two-dimensional coordinate is $\vec{r}=(x,y)$). An electric field E
is applied along the z-axis and electrons are confined in a
triangular potential $V(z)$, and it is assumed that only the lowest
subband along the z-axis is occupied.  Thus in our model quantum
dots are effectively  quasi-two dimensional. The Rashba constant
$c_R$ changes when the electric field E is varied. The potential
$V_p(\vec{r})=\epsilon' y$ perturbs the two dimensional harmonic
potential with the strengths  $\omega_x$ and $\omega_y $. This
potential can be realized by applying a constant electric field
along the y-axis and its strength $\epsilon'$ is controlled by the
magnitude of the applied electric field along the y-axis. The
crucial point about $U(\vec{r})$ and $V(z)$ is that they can be
changed {\it electrically}, which provides a means to control
coherently electron spins. The Rashba spin orbit term\cite{ras} is
$h_\mathrm{R}$ with Pauli spin matrices $\sigma_{x,y}$ and a
momentum operator $k_x=\frac{1}{i}\frac{d}{dx}$(similarly with
$k_y$). The Dresselhaus term can be also included, but since it does
not change results qualitatively we {\it omit} it here. The
Hamiltonian, Eq.(\ref{eq:singleHam}), represents a simple model of
the physical system, but it has all the correct symmetries.  It is
invariant under time reversal operation and each eigenenergy is
doubly degenerate. The Hamiltonian is  not invariant under
two-dimensional inversion operation $\vec{r}\rightarrow -\vec{r}$
since the Rashba spin-orbit term breaks inversion symmetry. In order
to build up many-body wavefunctions we need to first construct
single electron eigenstates. Each of these wavefunctions consists of
the spin-up and down components:
\begin{eqnarray}
|\phi\rangle= \left(
\begin{array}{c}
F_{\uparrow}(\vec{r}) \\
F_{\downarrow}(\vec{r})
\end{array}
\right)= \left(
\begin{array}{c}
\sum_{m n}c_{m n\uparrow }\langle \vec{r}|m n \uparrow \rangle \\
\sum_{m' n'}c_{m' n'\downarrow }\langle \vec{r}|m' n' \downarrow
\rangle
\end{array}
\right), \label{eq:single-eigen}
\end{eqnarray}
where $|m n  \rangle$ are eigenstates of two-dimensional harmonic
oscillators.

\section{Breaking of inversion symmetry, correlations, and matrix Berry phase}

 When the lateral potential has  two-dimensional inversion symmetry
the effect of many-body correlations will not produce  a finite
value of the matrix Berry phase. This can be shown to be an exact
result. (See Appendix B). It follows from the transformation
properties of the wavefunctions under inversion operation. It should
be stressed that although the lateral potential has inversion
symmetry the total Hamiltonian does not.
 When the  inversion symmetry of $U(\vec{r})$
is broken many-body correlations will induce a finite  value of the
matrix Berry phase. It is not possible to compute this effect
exactly unlike the case when $U(\vec{r})$ has inversion symmetry. In
a previous work a truncated single electron $4\times4$ Hamiltonian
matrix  was used\cite{yang1}. However, many-body  states built from
these approximate single electron wavefunctions do not adequately
describe many-body correlation effects. In this paper we  find
improved single electron wavefunctions, and use them to build
many-body wavefunctions. Here we will compute  the degenerate
groundstates approximately by using a finite number of Slater
determinant basis states, i.e., using $M=4$ in
Eq.(\ref{eq:groundstates}). This approximation should be valid as
long as the single electron energy spacing is larger than or
comparable to the characteristic Coulomb energy scale.

\subsection{Single electron Hamiltonian matrix}

We employ an  improved approximation of a $6\times6$ truncated
single electron Hamiltonian matrix, whose eigenvectors can be
written, according to Eq.(\ref{eq:single-eigen}), as
\begin{eqnarray}
| \phi \rangle
&\cong& c_{0,0,\uparrow } |00\uparrow\textgreater+c_{0,1,\uparrow } |0,1,\uparrow\textgreater+c_{0,2,\uparrow } |0,2,\uparrow\textgreater\nonumber \\
&+&c_{0,0,\downarrow } |0,0,\downarrow\textgreater+c_{0,1,\downarrow } |0,1,\downarrow\textgreater+c_{0,2,\downarrow } |0,2,\downarrow\textgreater. \nonumber \\
\end{eqnarray}
These expansion coefficients $c_{m n\sigma}$ of the basis states
$|mn\sigma\rangle$ are eigenvectors of the $6\times6$ Hamiltonian
matrix

%\begin{widetext}
\begin{eqnarray}
\ \ \ \ \ \ \ \ \fl H_{single}=\left(\begin{array}{cccccc}
E_{0} & E_{P} & 0 & 0 & -iE_{R} & 0 \\
E_{P} & E_{1} & \sqrt{2}E_{P} & iE_{R} & 0 & -\sqrt{2}iE_{R} \\
0 & \sqrt{2}E_{P} & E_{2} & 0 & \sqrt{2}iE_{R} & 0 \\
0 & -iE_{R} & 0 & E_{0} & E_{P} & 0 \\
iE_{R} & 0 & -\sqrt{2}iE_{R} &E_{P} & E_{1} & \sqrt{2}E_{P} \\
0 & \sqrt{2}iE_{R} & 0 & 0 & \sqrt{2}E_{P} & E_{2}
\end{array} \right).
\label{eq:single-ham}
\end{eqnarray}
The basis vectors of this Hamiltonian matrix are ordered as
$|00\uparrow\rangle$, $|01\uparrow\rangle$, $|02\uparrow\rangle$,
etc. To reduce the number of independent external parameters we can
set the ratio between the harmonic frequencies be a constant, for
example, $\omega_x=3\omega_y$. Then  the  energies of the
two-dimensional harmonic oscillators are $E_0=2\hbar\omega_y$,
$E_1=3\hbar\omega_y$, and $E_2=4\hbar\omega_y$.  The eigenvalues of
this matrix are
\begin{eqnarray}
X_{1}&=& \frac{3}{2} E_{0}-\frac{1}{3} \varepsilon \cos \left(\frac{\theta }{3}\right),\nonumber\\
X_{2}&=& \frac{3}{2} E_{0}+\frac{1}{6} \varepsilon \left(\cos
\left(\frac{\theta
   }{3}\right)-\sqrt{3} \sin \left(\frac{\theta }{3}\right)\right),\nonumber\\
X_{3}&=& \frac{3}{2} E_{0}+\frac{1}{6} \varepsilon \left(\cos
\left(\frac{\theta
   }{3}\right)+\sqrt{3} \sin \left(\frac{\theta }{3}\right)\right),
   \label{eq:AnalEigen}
\end{eqnarray}
where
\begin{eqnarray}
\varepsilon &=&\sqrt{3} \sqrt{E_{0}^2+12 \delta},\
\delta=E_{P}^2+E_{R}^2,\nonumber\\
A&=&-18E_0 \left(E_P^2+E_R^2\right),\nonumber\\
B&=&\frac{1}{6} \sqrt{108 \left(E_0^2+12 \delta\right)^3-11664 E_0^2 \delta^2},\nonumber\\
\cos \theta&=&\frac{A}{\sqrt{A^2+B^2}}, \ \sin
\theta=\frac{B}{\sqrt{A^2+B^2}}.
\end{eqnarray}
Because of time reversal symmetry each eigenenergy is doubly
degenerate. The doubly degenerate wavefunctions of the energy shells
with eigenenergies $X_1$, $X_2$, $X_3$ are denoted by
$(\phi_1,\phi_2)$, $(\phi_3,\phi_4)$, and $(\phi_5,\phi_6)$,
respectively. There is some arbitrariness in choosing these
eigenstates since new eigenstates may be obtained by applying
unitary transformation to the old set in each degenerate energy
shell\cite{Wil}. We  choose the  expansion coefficients of the
first, second, and third  pairs degenerate eigenstates of
Eq.(\ref{eq:single-ham}) as
\begin{eqnarray}
\vec{c}(1)&=&\frac{1}{\sqrt{N_{1}}}\left( \alpha_{1},\beta_{1},\gamma_{1},\delta_{1},0,1\right),\nonumber\\
\vec{c}(2)&=&\frac{1}{\sqrt{N_{1}}}\left(-\delta^*_{1},0,-1,\alpha^*_{1},\beta^*_{1},\gamma^*_{1}\right),
\label{eq:EigenSet1}
\end{eqnarray}
\begin{eqnarray}
\vec{c}(3)&=&\frac{1}{\sqrt{N_{2}}}\left( \alpha_{2},\beta_{2},\gamma_{2},\delta_{2},0,1\right),\nonumber\\
\vec{c}(4)&=&\frac{1}{\sqrt{N_{2}}}\left(-\delta^*_{2},0,-1,\alpha^*_{2},\beta^*_{2},\gamma^*_{2}\right),
\label{eq:EigenSet2}
\end{eqnarray}
and
\begin{eqnarray}
\vec{c}(5)&=&\frac{1}{\sqrt{N_{3}}}\left( \alpha_{3},\beta_{3},\gamma_{3},\delta_{3},0,1\right),\nonumber\\
\vec{c}(6)&=&\frac{1}{\sqrt{N_{3}}}\left(-\delta^*_{3},0,-1,\alpha^*_{3},\beta^*_{3},\gamma^*_{3}\right).
\label{eq:EigenSet3}
\end{eqnarray}
Here $\vec{c}(p)$ denotes the  expansion coefficients $\{c_{m
n\sigma}(p)\}$ of p'th eigenstate. The quantities $\alpha_p$,
$\beta_p$, $\gamma_p$,  $\delta_p$  are too complicated and lengthy
to give here, however, they are all purely real or imaginary. This
choice of the eigenstates simplifies the calculation of the
many-body NAVPs. When $U(\vec{r})$ has inversion symmetry, i.e.,
$E_{P}=0$, then $\alpha_{i}$ and $\gamma_{i}$ are zero.

\subsection{Many-body Hamiltonian matrix}

Using the previous results for single-electron wavefunctions we
construct many-body groundstates. Let us assume there are three
electrons in the dot.  In order to calculate the many-electron NAVPs
we need to compute the expansion coefficients $c_i$  and $d_i$ of
Eq.(\ref{eq:groundstates}). We include three single electron energy
shells, each with double degeneracy. In our approximation  we
truncate the number of  Slater determinant wavefunctions  to  four
with the lowest total confinement  energies. They are
\begin{eqnarray}
|\Psi_1\rangle&=&a_3^+a_2^+a_1^+|0\rangle,\nonumber\\
|\overline{\Psi}_1\rangle&=&a_4^+a_2^+a_1^+|0\rangle= |\Psi_2\rangle, \nonumber\\
|\Psi_3\rangle&=&a_5^+a_2^+a_1^+|0\rangle,\nonumber\\
|\overline{\Psi}_3\rangle&=&a_6^+a_2^+a_1^+|0\rangle=
|\Psi_4\rangle,
\end{eqnarray}
where $a_i^+$ creates an electron in the $i$'th single electron
eigenstate $|\phi_i\rangle$, given in
Eqs.(\ref{eq:EigenSet1})-(\ref{eq:EigenSet3}). The vacuum state is
$|0\rangle$. The truncated many-body Hamiltonian matrix is
\begin{eqnarray}
H=\left( \begin{array}{cccc}
E_{A} & 0 & a & b \\
0 & E_{A} & -b^{*} & a^{*} \\
a^{*} & -b & E_{B} & 0 \\
b^{*} & a & 0 & E_{B}
\end{array}\right),
\label{eq:Ham-matrix}
\end{eqnarray}
where  the matrix elements are
\begin{eqnarray}
\langle \Psi_{1}|H|\Psi_{1}\rangle&=&E_{A},\nonumber\\
\langle \Psi_{3}|H|\Psi_{3}\rangle&=&E_{B},\nonumber\\
\langle \Psi_{1}|V|\Psi_{3}\rangle&=&a ,\nonumber\\
\langle \Psi_{1}|V|\Psi_{4}\rangle&=& b, \nonumber\\
\langle\Psi_{1}|V|\Psi_{2}\rangle&=& 0\nonumber\\
\langle\Psi_{3}|V|\Psi_{4}\rangle&=& 0.
\end{eqnarray}
The matrix elements $a$ and $b$ contain Hartree and exchange
contributions.
All the quantities $E_A$, $E_B$, $a$, $b$   change when the adiabatic parameters change
because they depend on the single electron wavefunctions,
Eqs.(\ref{eq:EigenSet1})-(\ref{eq:EigenSet3}), that are functions of the adiabatic parameters.
The groundstate eigenenergy is doubly degenerate with
the value
\begin{eqnarray}
E_G&=&\frac{1}{2} \left(E_A+E_B-D^{\frac{1}{2}}\right),
\end{eqnarray}
where
\begin{eqnarray}
D&=&(E_{A}-E_{B})^2+4 (|a|^2+ |b|^2).
\end{eqnarray}
One of the doubly degenerate groundstates  has the expansion
coefficients
\begin{eqnarray}
(c_1,c_2,c_3,c_4)= \frac{1}{\sqrt{N_1}}\left(-bK ,-a^{*}K
,0,1\right), \label{eq:Cof1}
\end{eqnarray}
where
\begin{eqnarray}
K=\frac{(-E_{A} +E_{B} +D^{\frac{1}{2}})}{2
\left(|a|^2+|b|^2\right)}. \label{eq:Element}
\end{eqnarray}
The other degenerate state is obtained by taking time reversal of
this state
\begin{eqnarray}
(d_1,d_2,d_3,d_4)= \frac{1}{\sqrt{N_1}}\left(aK ,-b^*K ,-1,0\right).
\label{eq:Cof2}
\end{eqnarray}
Since $c_{mn\sigma}(p)$ of
Eqs.(\ref{eq:EigenSet1})-(\ref{eq:EigenSet3}) are purely real or
imaginary it  follows from the expression for the Coulomb matrix
elements, Eq.(\ref{eq:apB1}), that $a$ is real and $b$ is imaginary.
Then from Eq.(\ref{eq:Element}) we also see that $K$ is real. This
implies that the expansion coefficients $c_i$ and  $d_i$ are always
purely real or imaginary.

When inversion symmetry {\it present} in $U(\vec{r})$ the results
given in Eqs.(\ref{eq:Cof1}) and (\ref{eq:Cof2}) simplify since
$b=0$, which follows from
\begin{eqnarray}
b=\langle \Psi_{1}|H|\Psi_{4}\rangle&=&\langle 13|v|16 \rangle -\langle 13|v|16 \rangle  \nonumber\\
&+& \langle 23|v|26 \rangle-\langle 23|v|62 \rangle=0,
\end{eqnarray}
where $v$ is the Coulomb interactions between two electrons. (This
is because, as can be seen from
Eqs.(\ref{eq:ShellA})-(\ref{eq:ShellC}), $\phi_3(r)^*\phi_6(r)$,
$\phi_1(r)^*\phi_6(r)$, and $\phi_3(r)^*\phi_2(r) $ are odd
functions of $r$). In this case the expansion coefficients of the
doubly degenerate many-body groundstates are given by
\begin{eqnarray}
(c_1,c_2,c_3,c_4)=\frac{1}{\sqrt{N_1}}\left(0
,-aK,0,1\right)\nonumber\\
\label{eq:CofA}
\end{eqnarray}
and
\begin{eqnarray}
(d_1,d_2,d_3,d_4)=\frac{1}{\sqrt{N_1}}\left(aK ,0
,-1,0\right).\nonumber\\
\label{eq:CofB}
\end{eqnarray}
Note that $c_2^*d_3$ may not be zero, but
$(B_k)_{2,3}=(a_{k})_{4,5}=0$ from  Eq.(\ref{eq:single_inter}).

\subsection{Correlation and matrix Berry phase }

We are now ready to calculate the matrix Berry phase.
The strengths of the distortion potential and Rashba constant are
\begin{eqnarray}
E_P=\epsilon'\ell_y/\sqrt{2}\  \textrm{and}\  E_R=c_R/\sqrt{2}\ell_y.
\end{eqnarray}
We choose these parameters as the adiabatic parameters:
$\lambda_1=E_P$ and $\lambda_2=E_R$ (The single electron Hamiltonian
depends on them, see Eq.(\ref{eq:single-ham}). As explained in Sec.2
these parameters can be controlled electrically). The adiabatic path
is elliptic
\begin{eqnarray}
&&(\lambda_1(t),\lambda_2(t))=(E_R(t),E_P(t))= \nonumber \\
&&(E_{R,c}+\Delta E_R \cos(\omega t),E_{P,c}+\Delta E_P \sin(\omega
t)). \label{closed_loop}
\end{eqnarray}
We use  the parameters $E_{R,c}= 0.5 E_0$, $E_{P,c}=0.3E_0$, $\Delta
E_R= 0.35 E_0$, $\Delta E_P= 0.21 E_0 $, and $\omega=E_0 /10$. The
following steps are implemented consecutively in the computing the
matrix Berry phase:
\begin{enumerate}
\item[(a)]
Single electron eigenvectors are evaluated numerically  from
Eqs.(\ref{eq:EigenSet1})-(\ref{eq:EigenSet3}).
\item[(b)]
Coulomb matrix elements are computed using Eqs.(\ref{eq:Coulomb}).
\item[(c)]
Many electron eigenvectors are evaluated from  Eqs.(\ref{eq:Cof1})
and (\ref{eq:Cof2}).
\item[(d)]
The many-electron NAVPs , given by  Eqs.(\ref{eq:Diagonal}) and
(\ref{eq:OffDiagonal}), are evaluated on the various point on the
closed adiabatic path in the parameter space. For this purpose the
expansion coefficients $d_{i}$ are  differentiated numerically. Also
we use that the diagonal elements of the NAVPs between Slater
determinants, Eq.(\ref{eq:Slater}) are zero
\begin{eqnarray}
(B_k)_{i,i}=\sum_{p\in occ.}(a_{k})_{p,p}=0,
\end{eqnarray}
where the sum over $p$ indicates a sum over single electron states
that appear in the Slater determinant state $|\Psi_i\rangle$. This
follows from the fact that $c_{mn}(p)$ is always purely real or
imaginary, see Eqs.(\ref{eq:EigenSet1})-(\ref{eq:EigenSet3}).  Thus
$(a_{k})_{p,p}=\frac{i}{2}\frac{d}{d\lambda_k}[\sum_{mn}c^*_{mn}(p)
c_{mn}(p)]=0$ for each $p$. (Note that the harmonic oscillator
states $|mn\rangle$ do not depend on $\lambda_k$).  In addition
it follows from the orthonormalization  $\langle \Psi_i|\Psi_j\rangle=\delta_{ij}$ and Eq.(\ref{eq:Slater}) that
\begin{eqnarray}
(B_k)_{i,j}=(B_k)^{*}_{j,i} \ \textrm{for $i\neq j$}.
\end{eqnarray}
\end{enumerate}

\begin{figure}[!hbt]
\begin{center}
\includegraphics[width = 0.5 \textwidth]{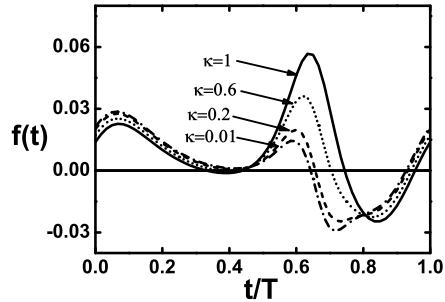}
\caption{$f(t)$ as a function of $t/T$ for $\kappa = 0.01, 0.2, 0.6,
1$. } \label{expo}
\end{center}
\end{figure}

\begin{figure}[!hbt]
\begin{center}
\includegraphics[width = 0.5 \textwidth]{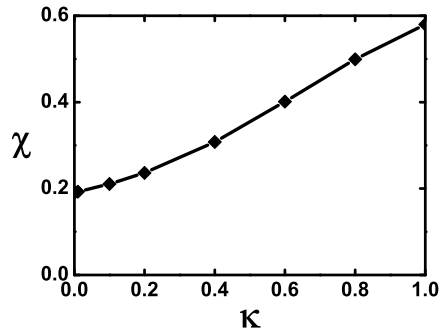}
\caption{The dependence of $\chi$ on  $\kappa $, where $\chi$
characterize the $2\times2$ matrix Berry phase and $\kappa $ is the
ratio between the Coulomb energy scale and single electron levels
spacing. $\kappa $ measures the strength of quantum fluctuations.  }
\label{chi}
\end{center}
\end{figure}

Using these results we find that the NAVPs, Eqs.(\ref{eq:Diagonal})
and (\ref{eq:OffDiagonal}), are off-diagonal:
\begin{eqnarray}A_{1}=
\left( \begin{array}{cc}
0 & P \\
P & 0 \\
\end{array}\right),\
A_{2}= \left( \begin{array}{cc}
0 & Q \\
Q & 0 \\
\end{array}\right)
\end{eqnarray}
with $P,Q$  real. The time dependent Schr\"{o}dinger equation for
the expansion coefficients $C_1$ and $C_2$,  given by
Eq.(\ref{eq:time_Schrod}),  can then be written as
\begin{eqnarray}
 \frac{dC_{1}}{dt}=if(t)C_{2}, \;\; \frac{dC_{2}}{dt}=if(t)C_{1},
\end{eqnarray}
where the function
\begin{eqnarray}
f(t)=P\frac{d\lambda_{1}}{dt}+Q\frac{d\lambda_{2}}{dt}.
\end{eqnarray}
For each  point $(\lambda_1(t),\lambda_2(t))$ on the adiabatic path
the off-diagonal elements of the NAVPs are evaluated numerically, as
described in steps (a), (b), (c), and (d) above. As the strength of
the Coulomb interaction $\kappa=\frac{e^2}{\epsilon \ell_y}/E_0$
increases $f(t)$ varies more significantly, see Fig.\ref{expo}. The
calculated  matrix Berry phase for $\kappa \le 1$ is given as
follows\cite{com1}
\begin{eqnarray}
\left(\begin{array}{c} C_1(T) \\ C_2(T)
\end{array}\right)
= \left( \begin{array}{cc}
\cos\chi & i\sin\chi \\
i\sin\chi & \cos\chi \\
\end{array}\right)
\left( \begin{array}{c} C_1(0) \\ C_2(0)
\end{array}\right),
\label{eq:matrixBerry}
\end{eqnarray}
where $T$ is the period of the adiabatic cycle. The parameter
$\chi=\int^{T}_{0}f(t)dt$ of Eq.(\ref{eq:matrixBerry}) is shown in
Fig.\ref{chi}. For $\kappa>1$ Slater determinant states with higher
total confinement energies than those four we have used  need to be
included in the many-body basis set. This also implies that   single
electron states with higher energies than those six we have used
must be included.

As a check on the correctness of our numerical procedures we have
verified numerically that  the value of $\cos(\chi)$ is independent
of the choice of the set of degenerate groundstates. (Although the
elements of the matrix Berry phase depend on the the choice of the
basis states the trace of it  is independent of the basis states).
By taking the limit of vanishing the strength of Coulomb interaction
$\kappa \rightarrow 0$ in Eqs.(\ref{eq:Cof1}) and (\ref{eq:Cof2}) we
see that the degenerate groundstates are $(c_1,c_2,c_3,c_4)=
\left(\alpha ,\beta ,0,0\right)$ and $(d_1,d_2,d_3,d_4)=
\left(-\beta^* ,\alpha^* ,0,0\right)$ (This limit is somewhat
delicate since $N_1$ and $K$ diverge).  We can also use another
possible set for degenerate groundstates $(c_1,c_2,c_3,c_4)= \left(1
,0 ,0,0\right)$ and $(d_1,d_2,d_3,d_4)= \left(0 ,1 ,0,0\right)$. The
elements of the NAVPs with respect to these new degenerate
groundstates are $(\tilde{A}_k)_{11}=(\tilde{B}_k)_{11}=0$,
$(\tilde{A}_k)_{12}=(\tilde{B}_k)_{12}=(a_k)_{34}$. However, we find
that the value of $\cos(\chi)$ is the same in these different sets
of ground states.

Using the computed matrix Berry phase, Eq.(\ref{eq:matrixBerry}), we
now evaluate the single electron occupation numbers, which can be
measured in tunneling experiments. Combining
Eqs.(\ref{eq:time-dep-state}) and (\ref{eq:groundstates}) we find
that the many-body groundstate at each time instant is given by
\begin{eqnarray}
|\Psi\rangle=C_1\sum_ic_i|\Psi_i\rangle+C_2\sum_id_i|\Psi_i\rangle.
\end{eqnarray}
The probabilities  that a single-electron eigenstate $p$ is occupied
at t=0 and T are, respectively,
\begin{eqnarray}
f_p(0)&=&\sum_i|(C_1(0)c_i(0)+C_2(0) d_i(0))|^2\theta_{ip}\nonumber\\
f_p(T)&=&\sum_i|(C_1(T)c_i(0)+C_2(T) d_i(0))|^2\theta_{ip},
\label{eq:occ}
\end{eqnarray}
where $f_p=\langle\Psi|a_p^+a_p|\Psi\rangle$, $c_i(T)=c_i(0)$, and
$d_i(T)=d_i(0)$ ($c_i$   and $d_i$ are given in Eqs. (\ref{eq:Cof1})
and (\ref{eq:Cof2})). If the single-electron eigenstate $p$  is
occupied (unoccupied) in the Slater determinant state
$|\Psi_i\rangle$ we define   $\theta_{ip}=1(0)$. At $\kappa=1$  we
find for the occupation number of the third single electron level
$f_3(0)=0.0266$ and  $f_3(T)=0.4494$. For the fourth single electron
level we find $f_4(0)=0.9492$ and  $f_4(T)=0.5264$. The difference
between $f_p(0)$ and $f_p(T)$ reflects the presence of a matrix
Berry phase. It would be interesting to measure these differences in
the single electron occupation numbers before and after an adiabatic
cycle.

\section{Discussions}

The Hamiltonian of II-VI and III-V n-type semiconductor quantum dots
with spin-orbit terms are not invariant under two-dimensional
inversion operator. Despite this, whether the lateral confinement
potential is or is not invariant has important consequences on the
matrix Berry phase. Our investigation shows that many-body
correlation effects do not generate a matrix Berry phase  when the
confinement potential is invariant under two-dimensional parity
operation. This is an exact result. It  holds despite that  the {\it
inter-shell} single electron NAVPs couple different single electron
energy levels. However, when the confinement potential is not
invariant under parity operation our approximate calculation
indicates that correlations can affect the matrix Berry phase
significantly.

Our results can be tested  experimentally in self-assembled dots
with wetting layers\cite{pet,War}  or in gated n-type semiconductor
dots\cite{Kat}. These quantum dots have several attractive features:
The lateral shape of the  dot can be  distorted electrically to
induce breaking of  two-dimensional inversion symmetry.  Moreover,
the electron number can be varied from one to several electrons.
These electric means for control provide  excellent opportunities to
test systematically the effect of many-body correlations. We have
investigated  quantitatively how quantum fluctuations affect the
matrix Berry phase when the strength of Coulomb interaction is
smaller or comparable to the single electron level spacing. In
self-assembled  dots  the characteristic scale of the single
electron level spacing is $10-40$meV, which is larger or comparable
to the Coulomb energy scale of $10$meV. However, in gated
semiconductor quantum dots\cite{Kat} the characteristic scale of the
single electron level spacing is a few meV, which is smaller than
the Coulomb energy scale. In order to obtain accurate results for
these dots one needs to include a large number of  Slater
determinant basis states and single electron states. Nonetheless,
even for these systems the matrix Berry phase should be absent when
two-dimensional inversion symmetry is present, which should be
experimentally testable. It should be noted that the matrix Berry
phase depends on the geometric properties of an adiabatic
path\cite{yang5}.

\ack This work was supported by The Second Brain Korea 21 Project
and by grant No.C00275(I00410) from Korea Research Foundation.

\appendix
\section{Single electron states and  lateral inversion  symmetry}

When the lateral confinement potential $U(\vec{r})$ has inversion
symmetry the single electron eigenstates  simplify. For a given
2-fold degenerate energy shell  we  choose \cite{yang3} one of the
eigenstates  as
\begin{eqnarray}
|\phi\rangle= \left(
\begin{array}{c}
F_{o}(\vec{r}) \\
F_{e}(\vec{r})
\end{array}
\right), or \, |\phi\rangle= \left(
\begin{array}{c}
F_{e}(\vec{r}) \\
F_{o}(\vec{r})
\end{array}
\right), \label{eq:deg}
\end{eqnarray}
where $F_{e}(\vec{r})$ and $F_{o}(\vec{r})$ are  even and  odd
functions of $\vec{r}$. Although $|\phi\rangle$ has even and odd
spinor components it is   not an eigenstate of the parity operator
since the Hamiltonian is not invariant under two-dimensional
inversion operation due to the Rashba term. (Note that any linear
combination of the two states of Eq.(\ref{eq:deg}) can also be
chosen as a  single electron basis state in the degenerate Hilbert
subspace).

We define a single- or many-electron wavefunction to have  a  A-type
property under parity operation if the spin-up part changes sign
under parity operation:
\begin{eqnarray}
\left(
\begin{array}{c}
F_{o}(\vec{r})  \\
F_{e}(\vec{r})
\end{array}
\right)\rightarrow \left(
\begin{array}{c}
-F_{o}^*(\vec{r})  \\
F_{e}^*(\vec{r})
\end{array}
\right) \ \; \Rightarrow   A-type.
\end{eqnarray}
A  wavefunction has a B-type  property under parity operation if the
spin-down part changes sign under parity operation:
\begin{eqnarray}
\left(
\begin{array}{c}
F_{e}(\vec{r})  \\
F_{o}(\vec{r})
\end{array}
\right)\rightarrow \left(
\begin{array}{c}
F_{e}^*(\vec{r})  \\
-F_{o}^*(\vec{r})
\end{array}
\right)\ \; \Rightarrow   B-type.
\end{eqnarray}
Each eigenstate $|\phi_p\rangle$ can be labeled by a subscript $p$.
When $p$ is  odd  the spin-up and -down components of the
wavefunction are, respectively,  odd and even functions of
$\vec{r}$. When  $p$ is  even   the odd and even properties are
reversed.

\begin{figure}[!hbt]
\begin{center}
\includegraphics[width = 0.4 \textwidth]{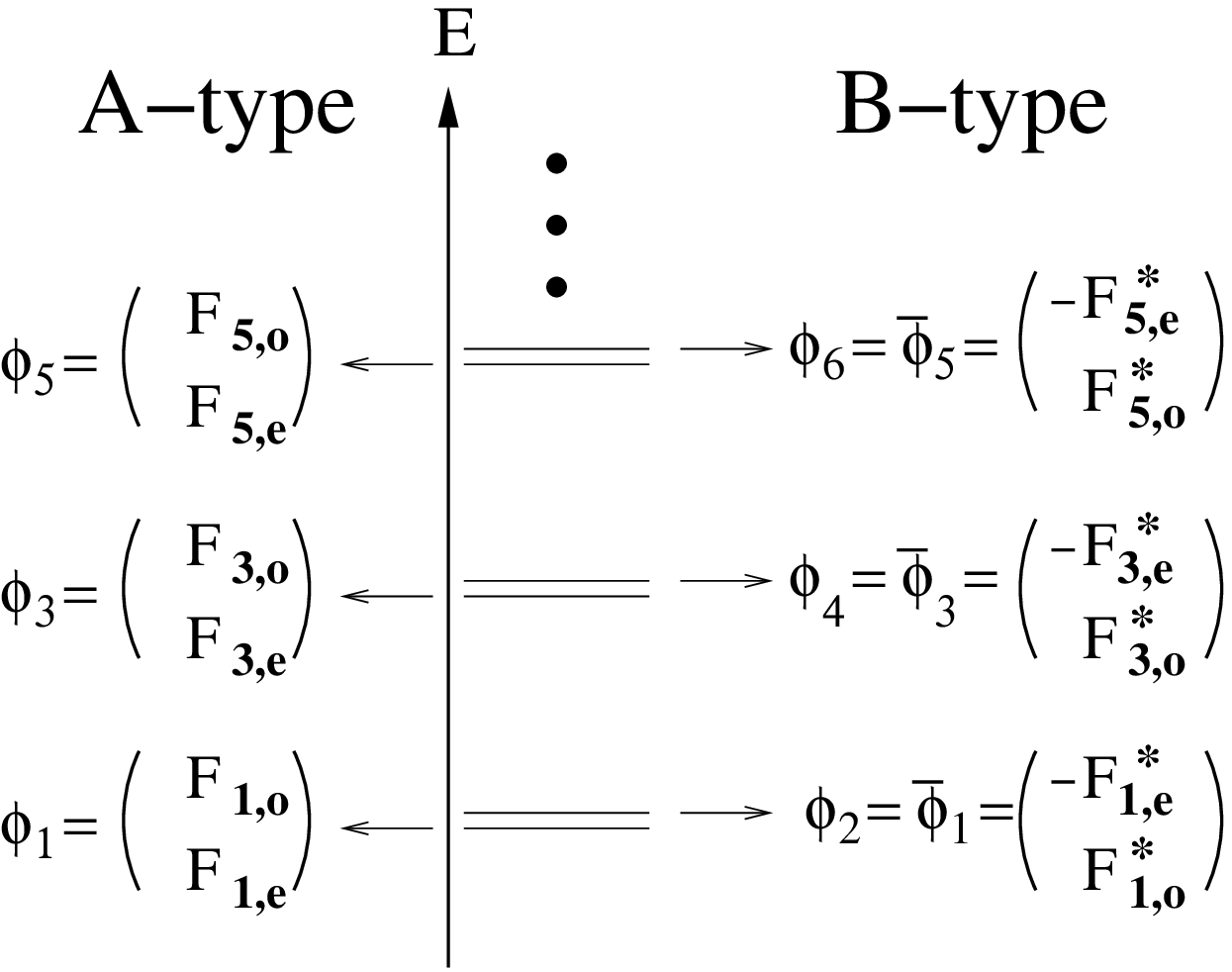}
\caption{ Each degenerate pair of eigenstates consists of  A- and
B-types.  These two types of eigenstates are time-reversed states of
each other.   As the subscript $i$ in $\phi_i$ increases the
transformation properties of $\phi_i$ alternate between A- and B-
types.} \label{fig:AB}
\end{center}
\end{figure}

In order to include many-electron physics we need to fix single
electron eigenstates of not only the first shell, but also of the
second,  third, and etc. energy shells. Here we choose them in the
following specific order
\begin{eqnarray}
|\phi_1\rangle= \left(
\begin{array}{c}
F_{1,o}(\vec{r}) \\
F_{1,e}(\vec{r})
\end{array}
\right), |\phi_2\rangle= \left(
\begin{array}{c}
-F_{1,e}^*(\vec{r})  \\
F_{1,o}^*(\vec{r})
\end{array}
\right), \label{eq:ShellA}
\end{eqnarray}
\begin{eqnarray}
|\phi_3\rangle= \left(
\begin{array}{c}
F_{3,o}(\vec{r}) \\
F_{3,e}(\vec{r})
\end{array}
\right), |\phi_4\rangle= \left(
\begin{array}{c}
-F_{3,e}^*(\vec{r})  \\
F_{3,o}^*(\vec{r})
\end{array}
\right), \label{eq:ShellB}
\end{eqnarray}
\begin{eqnarray}
|\phi_5\rangle= \left(
\begin{array}{c}
F_{5,o}(\vec{r}) \\
F_{5,e}(\vec{r})
\end{array}
\right), |\phi_6\rangle= \left(
\begin{array}{c}
-F_{5,e}^*(\vec{r})  \\
F_{5,o}^*(\vec{r})
\end{array}
\right),
\label{eq:ShellC}  \\
etc,\qquad\qquad\qquad\qquad \nonumber
\end{eqnarray}
Note that the wavefunctions of a degenerate pair are chosen to be
time-reversed states of each other. We have chosen  the single
electron wavefunctions $\phi_1, \phi_3,...$ to have A-type property,
and $\phi_2, \phi_4,...$ to have B-type  property under parity
operation, as shown in Fig.\ref{fig:AB}. This particular choice
simplifies the calculation of matrix Berry phases in the presence of
many-body correlation effects.   This corresponds to fixing a
convenient 'gauge', i.e., a single electron basis set.

The many-electron NAVPs contain single-electron NAVPs via
Eqs.(\ref{eq:Slater}) and (\ref{eq:inter}). So we need to understand
first the properties of  single-electron NAVPs. We can choose the
adiabatic parameters as $\lambda_1=2\hbar\omega_x$ and
$\lambda_2=\frac{c_R}{\sqrt{2}\ell_y}$, where the lengths are
$\ell_{x,y}=\sqrt{\hbar/m^*\omega_{x,y}}$. (The single electron
Hamiltonian depends on them, see Eq.(\ref{eq:singleHam})). The
adiabatic constant  $\lambda_1$ may be varied using the gate
potential of the dot and $\lambda_2$ may be  varied by changing the
electric field E along the z-axis. The single electron {\it intra-
shell} NAVP elements\cite{Wil} are $i\langle
\phi_{p}|\frac{\partial}{\partial\lambda_k
}|\overline{\phi}_{p}\rangle$, where $\phi_{p} $ and
$\overline{\phi}_{p}$ are degenerate single electron eigenstates.
The NAVP elements between A and B or of B and A states can be shown
to be zero\cite{yang3}. Since  $\phi_{p} $ and  $\overline{\phi}_{p}
$ are either of A and B or of B and A  the {\it intra-shell} NAVP
elements are zero. On the other hand, from the transformation
properties of the eigenstates, given in
Eqs.(\ref{eq:ShellA})-(\ref{eq:ShellC}), we can show that the single
electron {\it inter-shell} NAVP elements are
\begin{eqnarray}
(a_{k})_{p,q}\neq0 \  \textrm{if $p+q$  even},
\end{eqnarray}
and
\begin{eqnarray}
(a_{k})_{p,q}=0 \ \textrm{if $p+q$  odd}, \label{eq:single_inter}
\end{eqnarray}
where $\phi_{p} $ and $\phi_{q}$ belong to different energy shells.
Note that different single electron eigenstates can be   coupled
through $(a_{k})_{p,q}$ if $p+q$  is even. Thus the off-diagonal
many-electron NAVPs, Eq.(\ref{eq:OffDiagonal}), can  be written in
terms of  {\it non-zero} {\it inter-shell} single electron  NAVPs.
Nonetheless it is possible to show that many-electron matrix Berry
phase vanishes.

\begin{figure}[!hbt]
\begin{center}
\includegraphics[width = 0.45 \textwidth]{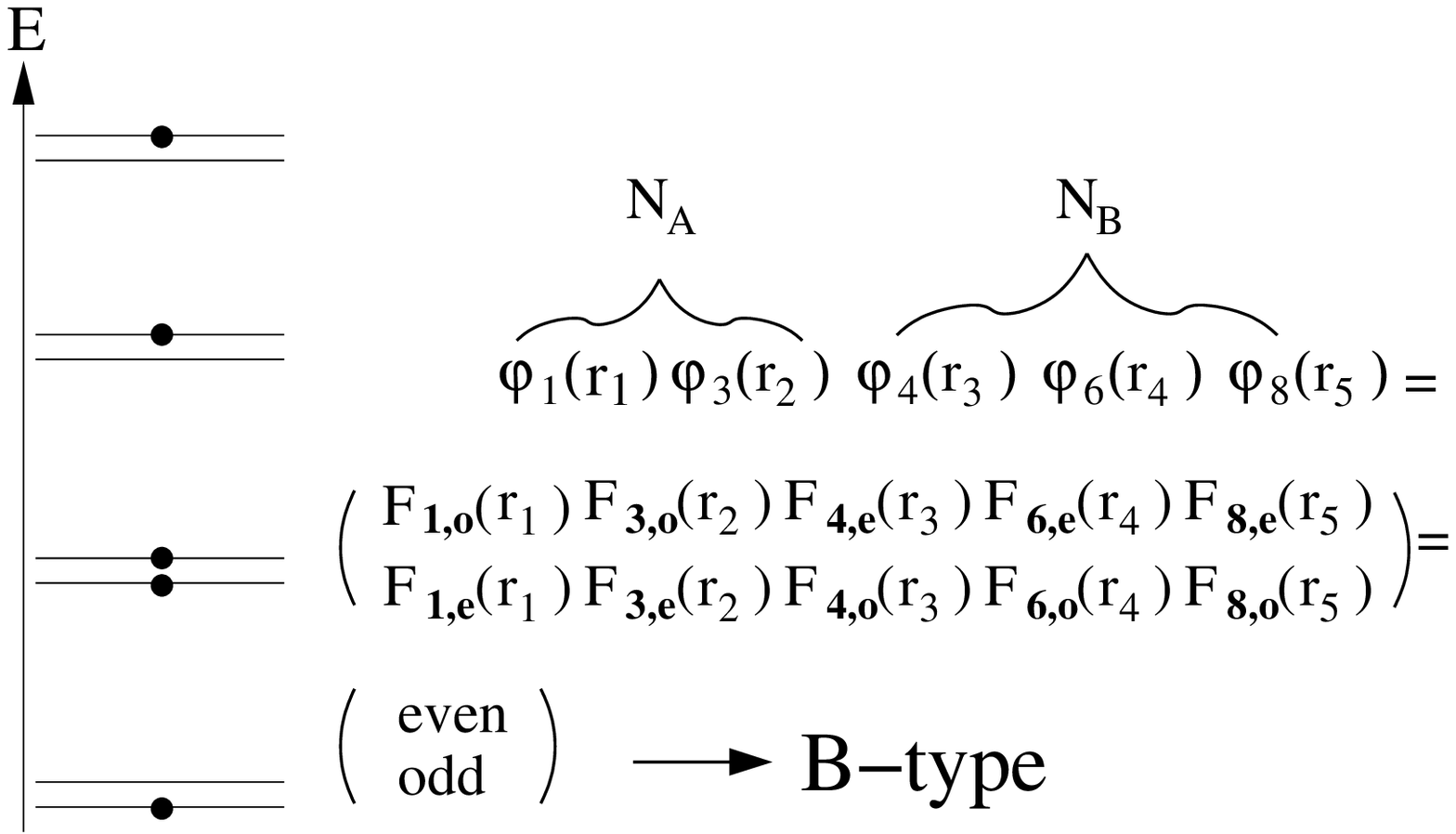}
\caption{ A B-type Slater determinant is a sum of  $N!$ terms.   One
of these terms  is  shown. } \label{fig:AB2}
\end{center}
\end{figure}

\section{Absence of matrix Berry phase and  lateral inversion symmetry}

Many electron states can be written as a linear combination of
Slater determinant states $|\Psi_i\rangle$. In the following we will
choose $|\Psi_1\rangle, |\Psi_3\rangle, ...$ as A-type Slater
determinant states, and $|\Psi_2\rangle=\hat{T}|\Psi_1\rangle,
|\Psi_4\rangle=\hat{T}|\Psi_3\rangle,...$ as  B-type Slater
determinant states. The time-reversed state of  A-type single
electron wavefunctions are of  B-type, and vice versa. The Slater
determinant states $|\Psi_i\rangle$ are chosen in the order of
increasing confinement  energy $\langle
\Psi_i|H_K+V_C|\Psi_i\rangle$. The total number of electrons
$N=N_A+N_B$ is odd with the number of  A-type single electron
wavefunctions  $N_A$ and of B-type  $N_B$. As explained in
Fig.\ref{fig:AB2}
 if  $N_B$ odd the Slater determinant  transforms like a B-type i.e., spin-down  part of the wavefunction changes sign.
On the other hand, if $N_A$ odd $|\Psi_i\rangle$ transforms like a
A-type, i.e., spin-up part of the wavefunction changes sign.
It can be shown that the NAVP between A and B Slater determinant states is zero
\begin{eqnarray}
( B_k)_{i,j}=i\langle
\Psi_{i}(A)|\frac{\partial}{\partial\lambda_k}|\Psi_{j}(B)\rangle=0.
\end{eqnarray}
This is because  NAVPs between A and B single electron  states are zero.

We find that a correlated degenerate groundstate, $|\Phi\rangle$ or
$|\overline{\Phi}\rangle$, is  either A- or B-type. This is because
the many-body Hamiltonian matrix element between  A-type and B-type
Slater determinant wavefunctions is zero, $\langle
\Psi_{i}(A)|H|\Psi_{j}(B)\rangle=0$: If $|\Phi\rangle$ is A-type and
$|\overline{\Phi}\rangle$ is B-type then
\begin{eqnarray}
|\Phi\rangle &=&c_1|\Psi_1(A)\rangle+c_3|\Psi_3(A)\rangle+...\nonumber\\
|\overline{\Phi}\rangle&=&d_2|\Psi_2(B)\rangle+d_4|\Psi_4(B)\rangle+...
\label{eq:AB}
\end{eqnarray}
We see from these results that, for a given index $i$, if an
expansion coefficient $c_i$ of one  degenerate groundstate is  zero
then the expansion coefficient $d_i$ of the other time-reversed
groundstate is non-zero, and vice versa.

The off-diagonal elements of the many-body NAVPs,
Eq.(\ref{eq:OffDiagonal}), are  zero. This can be shown as follows:
according to  Eq.(\ref{eq:AB}),  for each $i$, we have $c_i=0$ or
$d_i=0$, which implies that the first term of
Eq.(\ref{eq:OffDiagonal}) is $\sum_ic_i^*\frac{\partial
d_i}{\partial \lambda_{k}}=0$. From Eq.(\ref{eq:AB}) we see that
when $i\neq j $ and    $c^*_id_j$ is non-zero then  $\Psi_i$ and
$\Psi_j$ are of  A- and  B-types, respectively. But  this implies
$(B_k)_{i,j}=0$, and the product $c_i^*d_j( B_k)_{i,j}=0$. The
second term of Eq.(\ref{eq:OffDiagonal}) is thus
$\sum_{i,j}c_i^*d_j( B_k)_{i,j}=0$. An explicit example of this is
given below Eq.(\ref{eq:CofB}). There is thus a delicate interplay
between the many-body expansion coefficients $c_i^*d_j$ and the
elements of the NAVPs between Slater determinant states $(
B_k)_{i,j}$. Since $c_i^*\frac{\partial d_i}{\partial
\lambda_{k}}=0$ and $c_i^*d_j( B_k)_{i,j}=0$ the off-diagonal
elements of the many body NAVPs are zero: $(A_k)_{1,2}=0$. The
matrix Berry phase is thus {\it absent} for doubly degenerate
correlated states when inversion symmetry is present. This is true
at any level of approximation represented by the number of Slater
determinant states, $M$,  included in Eq.(\ref{eq:groundstates}).
Therefore, this is an {\it exact} result valid for $M\rightarrow
\infty$.

\section{Coulomb matrix elements}
The diagonal and off-diagonal matrix elements of $H$,
Eq.(\ref{eq:Ham-matrix}), depend on two-particle Coulomb matrix
elements between single electron eigenstates $p,q,r,s$, that are
given in Eqs.(\ref{eq:EigenSet1})-(\ref{eq:EigenSet3}),
\begin{eqnarray}
\fl \langle pq |v|rs\rangle&=& \sum_{\begin{array}{cccc}
\textrm{\scriptsize{$m_{p}$}},\textrm{\scriptsize{$m_{q}$}},\textrm{\scriptsize{$m_{r}$}},\textrm{\scriptsize{$m_{s}$}}, \\
\textrm{\scriptsize{$n_{p}$}},\textrm{\scriptsize{$n_{q}$}},\textrm{\scriptsize{$n_{r}$}},\textrm{\scriptsize{$n_{s}$}},\\
\textrm{\scriptsize{$\sigma_{p}$}},\textrm{\scriptsize{$\sigma_{q}$}},\textrm{\scriptsize{$\sigma_{r}$}},\textrm{\scriptsize{$\sigma_{s}$}}
\end{array}}
\delta_{\sigma_{p}\sigma_{r}}\delta_{\sigma_{q}\sigma_{s}}c^{\ast}_{m_{p}n_{p}}(p)c^{\ast}_{m_{q}n_{q}}(q)c_{m_{r}n_{r}}(r)
c_{m_{s}n_{s}}(s)\nonumber\\
&\times&  \langle m_{p}n_{p},m_{q}n_{q}|v|m_{r}n_{r},m_{s}n_{s}\rangle \nonumber \\
\label{eq:apB1}
\end{eqnarray}
where the  Coulomb matrix elements between eigenstates of
two-dimensional harmonic oscillator are
\begin{eqnarray}
\fl \langle m_{p}n_{p},m_{q}n_{q}|v|m_{r}n_{r},m_{s}n_{s}\rangle
=e^{2}&\int& d^{2}k \ \frac{1}{2\pi k }\ \langle m_{p}|e^{i k_{x}
x_{1}}|m_{r}\rangle \langle n_{p}|e^{i k_{y}
y_{1}}|n_{r}\rangle\nonumber\\
&\times&\langle m_{q}|e^{-i k_{x}
x_{2}}|m_{s}\rangle\langle n_{q}|e^{-i k_{y}
y_{2}}|n_{s}\rangle\nonumber\\
\label{eq:Coulomb}
\end{eqnarray}
with
\begin{eqnarray}
\langle m|e^{i k_{x} x}|m'\rangle =\left\{
\begin{array}{cc} (\frac{m' !}{m !})^{1/2} (\frac{i k_{x}l_{x}
}{\sqrt{2}})^{m-m'} e^{-\frac{k^{2}_{x}l^{2}_{x}}{4}}
L^{m-m'}_{m'}(\frac{k^{2}_{x}l^{2}_{x}}{2}) & (m' \leq m)\\ (\frac{m
!}{m' !})^{1/2} (-\frac{i k_{x}l_{x} }{\sqrt{2}})^{m'-m}
e^{-\frac{k^{2}_{x}l^{2}_{x}}{4}}
L^{m'-m}_{m}(\frac{k^{2}_{x}l^{2}_{x}}{2}) & (m \leq m')
\end{array}\right.\nonumber\\
\end{eqnarray}
and  Lagurre polynomials  $ L^{m'}_{m}(x)$. Similar expression can
be found  for $\langle n|e^{i k_{y} y}|n'\rangle$ with $\ell_y$
replacing $\ell_x$.
%\end{widetext}

%\end{references}


\begin{thebibliography}{10}
\bibitem{com1} Our matrix Berry phase is parameterized by a single parameter $\chi$ for the special choice of basis
vectors given in equations
(\ref{eq:EigenSet1})-(\ref{eq:EigenSet3}).
\end{thebibliography}
\end{document}